\documentclass[12pt]{iopart}
\usepackage[dvips]{graphicx}
\begin{document}
\title{Pulsar kicks by anisotropic neutrino emission from quark matter}
\author{I Sagert and J Schaffner-Bielich}
\address{Institut f\"ur Theoretische Physik, 
J. W. Goethe Universit\"at,
Max-von-Laue-Stra\ss{}e 1,
D-60438 Frankfurt,
Germany}
\eads{\mailto{sagert@astro.uni-frankfurt.de}, \mailto{schaffner@astro.uni-frankfurt.de}}
\begin{abstract}
We discuss an acceleration mechanism for pulsars out of their supernova remnants based on asymmetric neutrino emission from quark matter in the presence of a strong magnetic field. The polarized electron spin fixes the neutrino emission from the direct quark Urca process in one direction along the magnetic field. We calculate the magnetic field strength which is required to polarize the electron spin as well as the required initial proto-neutron star temperature for a successfull acceleration mechanism. In addition we discuss the neutrino mean free paths in quark as well as in neutron matter which turn out to be very small. Consequently, the high neutrino interaction rates will wash out the asymmetry in neutrino emission. As a possible solution to this problem we take into account effects from colour superconductivity.
\end{abstract}
\maketitle
\section{Introduction}
The birth of a neutron star in a supernova explosion releases the gravitational binding energy of about $10^{53}$ ergs which is mainly (to 99 \%) carried off by neutrinos. Within minutes after the explosion the new-born proto-neutron star cools down from temperatures of 50 MeV to $T<1$ MeV by the emission of neutrinos \cite{Pons99,Lattimer04}. During this Kelvin-Helmholtz cooling phase the density in the proto-neutron star can exceed several times normal nuclear density allowing for the creation of exotic matter such as hyperons, Bose condensates or strange quark matter \cite{Weber1}. Quark matter can be either present just in the core (a hybrid star) or entirely fill up the compact star except for a thin nuclear crust, a so-called strange star (for an overview of strange quark matter in compact stars see e.g. \cite{Schaffner-Bielich05}). For very high densities with quark chemical potentials larger than $\sim 400$ MeV quarks of all flavours and colours pair to the so-called colour-flavour locked phase (CFL), where charge neutrality is given by the strange quarks \cite{Ruester}. For temperatures larger than $\sim 10$ MeV R{\"u}ster et al. \cite{Ruester04} showed that the metallic CFL phase can appear where the electron chemical potential can be very large. The global properties of neutron stars and quark stars can be very similar, both have a radius of about 10 km and a mass of about two solar masses. Observable as pulsars they can possess very high magnetic fields up to  $10^{12}$ Gauss at the surface, for magnetars even up to $B\sim 10^{15}$ G. Observations show that many pulsars have very high space velocities between 100 km/s and 1600 km/s with the highest directly measured value for the pulsar B1508+55 of $1083^{+103}_{-90}$ km/s \cite{Brisken}. In addition, the Crab, the Vela pulsar and B0656+14 show an alignment between the velocity vector and the rotational axis, which is also seen for 25 pulsars by polarization measurements \cite{Johnston}. Since the discovery of the high neutron star space velocities, there has been a large number of suggestions to explain their origin ( see e.g. \cite{Wang,Lai,Blaschke,Janka}). A recent analysis by Ng and Romani \cite{Ng07} favour a magnetic-neutrino driven acceleration mechanism due to an magnetic field induced asymmetry of a neutrino driven kick with characteristic time scales for the anisotropy of 1-3 s. In \cite{Horowitz_Piekarewicz} as well as \cite{Horowitz_Li} the authors examined anisotropies in neutrino-momentum distribution in magnetized proto-neutron stars as an origin of pulsar kicks. However, it was shown that the neutrino anisotropy would be washed out in thermal and statistical equilibrium due to the large neutrino scattering rate \cite{Vilenkin}. The results from \cite{Ng07} give rise to the question whether an asymmetry in neutrino emission by the direct quark Urca process
$(d \rightarrow u + e^{-} + \bar{\nu_{\rm e}}$, $u + e^{-} \rightarrow \nu_{\rm e} + d)$ would be more successfull in accelerating the neutron star to high velocities. A strong magnetic field can align the electron spin opposite to the magnetic field direction. The electron spin polarization fixes the neutrino emission in one direction along the magnetic field and the neutrinos accelerate the neutron star. To check whether the neutrino acceleration can work we need to find the magnetic field strengths which are required to polarize the electrons. Furthermore the neutrino energy should be high enough to accelerate the neutron star and the neutrino mean free paths large enough to ensure free streaming neutrinos. 
\section{Polarization of the electrons - Landau levels}
\begin{figure}
{\centering \includegraphics[width=6cm, angle=270]{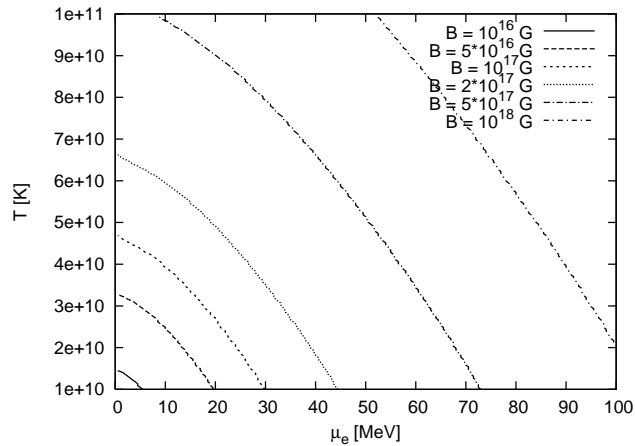} \par}
\caption{Constraints for the electron chemical potential $\mu_e$ and temperature $T$ for different magnetic field strengths $B$.}
\label{figure1}
\end{figure}

Electrons which are moving in a magnetic field stronger than $B_{\rm crit} \sim m_{\rm e} c^2/e\hbar \sim 4.4\cdot 10^{13}$ Gauss are situated in Landau levels perpendicular to the magnetic field axis. Their energy is quantized and can be written as $E_n^2  = m_{\rm e}^2 + p_{\rm z}^2 + 2eB\eta = m_{\rm e}^2 + p_{\rm z}^2 + 2eB(\nu + 1/2 + s)$, where $\eta$ is the Landau level number which can be expressed by it's quantum number $\nu$ and the electron spin $s=1/2$ or $s=-1/2$. For very high magnetic fields all electrons occupy the lowest Landau level with $\eta=0$. From the definition of the energy one sees that the lowest Landau level is populated by spin polarized electrons with $s=-1/2$. The degree of spin polarization $\chi$ can be expressed by the electron number density in the lowest Landau level $n_{\eta=0}$ over the total electron number density $n_{\rm total}$ as $\chi = n_{\eta=0}/n_{\rm total}$. For a sufficient acceleration we consider the largest possible value for the polarization, namely $\chi=1$. Figure \ref{figure1} shows the dependence of the full spin polarization on the electron chemical potential and the temperature for different magnetic fields. The fields strengths range from $B=10^{16}$ G to $10^{18}$ G where the possible values for $\mu_{\rm e}$ and $T$ increase with $B$. 
\section{Calculation of kick velocities}
The kick velocity can be calculated by integrating the stored thermal energy:
\begin{eqnarray}
v&=& \frac{4}{3} \pi R^3 \frac{\chi}{M_{\rm ns}}\int c_{\rm q} dT.
\label{equation1}
\end{eqnarray}
where $R$ is the neutrino emitting quark phase radius, $\chi$ the fraction of spin polarized electrons which will be set to one in the following and $M_{\rm ns}$ the total mass of the neutron star. The quark heat capacity $c_{\rm q} =9 \mu_{\rm q}^2 T (1-2\alpha_s/ \pi )$ gives the thermal energy stored in the medium and is therefore crucial for the kick velocity. The contribution from the electron heat capacity is small in comparison to $c_{\rm q}$ and can be neglected here. Considering a realistic value of $R\sim 10$ km we find from (\ref{equation1}) that velocities of 1000 km/s can be reached if the temperature is around $5\cdot 10^{10}$ K. For larger temperatures the quark phase radius can be smaller. These velocities are calculated assuming fully spin polarized electrons. For such temperatures we find from figure \ref{figure1} that the required magnetic fields are in the range of $\sim 10^{17}$ G to $10^{18}$ G depending on the electron chemical potential. These are of course very high values, however, the large magnetic field is supposed to be present in the quark phase and can be much smaller at the surface. 
\section{Neutrino mean free paths}
\begin{figure}
{\centering \includegraphics[width=6cm, angle=270]{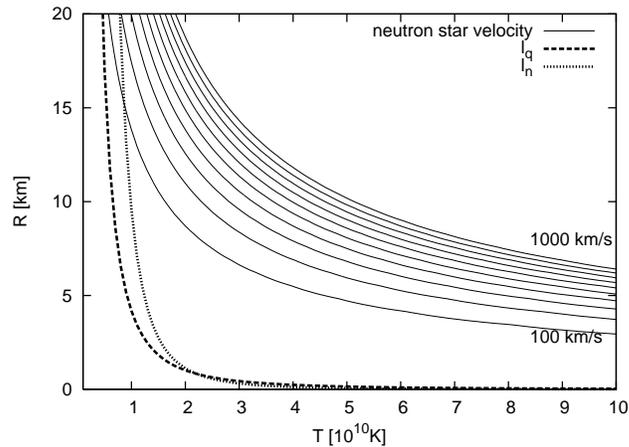} \par}
\caption{Pulsar kick velocities and neutrino mean free paths in dependence of the quark phase temperature for $\mu_{\rm q}=400$ MeV, $\mu_{\rm e}=10$ MeV, $M_{\rm ns}=1.4 M_\odot$ and $\chi=1$.}
\label{figure2}
\end{figure}

In regard to \cite{Vilenkin} we have to consider four neutrino interaction processes, namely scattering and absorption in quark matter as well as in neutron matter for a hybrid star. Taking calculations from \cite{Iwamoto81} we arrive at the neutrino mean free paths presented in table \ref{table1}.
\begin{table}
\begin{tabular}{|l||l||l|}
\hline
Medium & $E_\nu=3T, \alpha_{\rm s}=0.5$ &T=5MeV, $\mu_{\rm e}$=10MeV\\
\hline
\hline
Quark matter&$l_{\rm abs}\sim$25km$\left(\frac{T}{{\rm MeV}}\right)^{-2}\left(\frac{\mu_{\rm q}}{\rm  400 MeV}\right)^{-2}\left(\frac{\rm MeV}{\mu_{\rm e}}\right)$&$\sim$100m for $\mu_{\rm q}$=400MeV \\ 
Quark matter&$l_{\rm scat}\sim$92km$\left(\frac{\mu_{\rm q}}{\rm 400 MeV}\right)^{-2}\left(\frac{T}{\rm MeV}\right)^{-3}$&$\sim$800m for $\mu_{\rm q}$=400MeV\\
Neutron matter&$l_{\rm abs}\sim$17km$\left(\frac{T}{\rm MeV}\right)^{-3}\left(\frac{n_{\rm n}}{n_0} \right)^{-1/3}$&$\sim$150m for $n_{\rm n}=n_0$\\
Neutron matter&$l_{\rm scat}\sim$244km$\left(\frac{T}{\rm MeV}\right)^{-4}\left(\frac{n_{\rm n}}{n_0} \right)^{-2/3}$&$\sim$244m for $n_{\rm n}=n_0$\\
\hline
\end{tabular}
\caption{Neutrino mean free paths for absorption and scattering processes in quark matter as well as in neutron matter.}
\label{table1}
\end{table}
In figure \ref{figure2} we plot the neutron star velocities in dependence of the quark phase temperature and it's radius as well as the neutrino mean free paths defined as $1/l_{\rm i} = 1/l_{\rm abs}+1/l_{\rm scat}$ for quark and neutron matter. The lowest velocity of $v=100$ km/s is given by the first line and increases in steps of 100 km/s to $v=1000$ km/s. For a temperature of $T\geq 5\cdot 10^{10}$ K and an electron chemical potential of $\mu_{\rm e}=10$ MeV the neutrino mean free paths are very small. They increase for temperatures of $T \leq 10^{10}$ K, however for these small values it is impossible to reach high velocities within a realistic quark phase radius in the range of 10 km.
\section{Effects from colour superconductivity}
\begin{figure}
{\centering \includegraphics[width=6cm, angle=270]{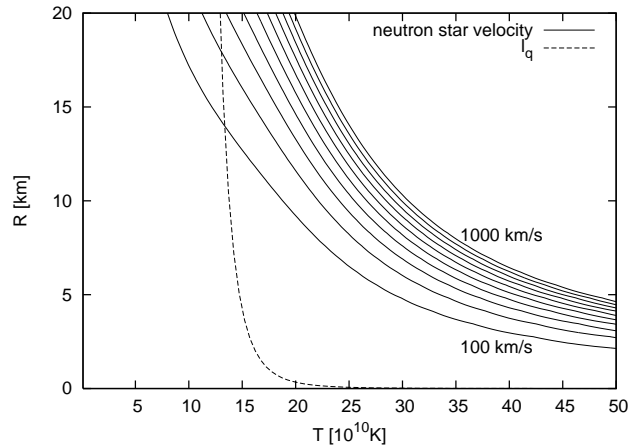} \par}
\caption{Pulsar kick velocities and neutrino mean free paths in dependence of the quark phase temperature in colour superconducting quark matter with $\Delta=100$ MeV, $\mu_{\rm q}=400$ MeV and $\mu_{\rm e}=100$ MeV.}
\label{figure3}
\end{figure}

A solution to the small neutrino mean free path in normal strange quark matter might be possible by taking into account effects of colour superconducting quark matter, where quarks pair with a pairing energy $\Delta$. The neutrino quark interactions are then suppressed exponentially by $e^{-\Delta  /T}$, for $T$ being smaller than a critical temperature $T_{\rm c}$. Unfortunately, the quark heat capacity is decreased by the same factor which will drastically lower the pulsar acceleration. Due to the smallness of $c_q$ we take into account the heat capacity of the electrons. The neutron star velocity can then be written as:
\begin{eqnarray}
v = \frac{2}{3} \pi R^3 \left(\left(\frac{\mu_{\rm e}^2}{2}+\frac{7\pi^2T^3}{60}\right)+A\mu_{\rm q}^2e^{-\Delta /T}T^2\right).
\end{eqnarray}
In figure \ref{figure3} we plot the new velocity curves as well as the neutrino mean free paths in quark matter. The consequences of the quark pairing energy is clearly seen. While the neutrino mean free paths are now much larger for temperatures in the range of 5 -10 MeV the neutron star velocity curves are shifted to higher temperatures. Consequently, it is again impossible to reach neutron star velocities for realistic quark phase radii of $\sim 10$ km.
\section{Conclusion}
We discussed an accelerating mechanism for pulsars based on anisotropic neutrino emission from quark matter due to a strong magnetic field during the proto-neutron star cooling stage. The asymmetry is generated by the electron spin polarization in the magnetic field. We found that for an initial temperature of $5\cdot 10^{10}$ K and a magnetic field in the range of $\sim 10^{17}$ G a velocity of 1000 km/s is reachable. Unfortunately, the high temperatures and densities cause the neutrinos to interact on their way to the surface of the neutron star, so that the neutrinos isotropize and the acceleration effect will be washed out. The presence of colour superconducting quark matter increases the neutrino mean free path but also decreases the quark heat capacity which is essential for the acceleration. Even when including the electron heat capacity with very high electron chemical potentials it is impossible to reach high pulsar velocities with a realistic quark phase radius of $\sim 10$ km. 
\section{References}
\bibliography{all}
\end{document}